\input epsf.tex
\input harvmac
\def\lsim{\,\,\raise2pt\hbox
   {$\mathop < \limits_{\raise0.5pt\hbox{$\sim$}}$}\,\, }
\def\gsim{\,\,\raise2pt\hbox
   {$\mathop > \limits_{\raise0.5pt\hbox{$\sim$}}$}\,\, }

\def\NP#1{Nucl.\ Phys. {\bf B{#1}}}
\def\PL#1{Phys.\ Lett. {\bf B{#1}}}

\def\PRB#1{Phys.\ Rev. {\bf B{#1}}}
\def\PRD#1{Phys.\ Rev. {\bf D{#1}}}

\def\PRL#1{Phys.\ Rev.\ Lett. {\bf {#1}}}
\def\PTP#1{Prog.\ Theor.\ Phys. {\bf {#1}}}

\Title{KYUSHU-HET-26, SAGA-HE-91}
{\vbox{\centerline{Topological charge distribution}
\vskip0.2em\centerline{and }
\vskip0.2em\centerline{$CP^1$ model with $\theta$ term}}}
\centerline{Ahmed S. Hassan, Masahiro Imachi\footnote{*}
{ e-mail:imac1scp@mbox.nc.kyushu-u.ac.jp},
 Norimasa Tsuzuki\footnote{**}
{ e-mail:tsuz1scp@mbox.nc.kyushu-u.ac.jp}}
\centerline{and}
\centerline{ Hiroshi Yoneyama$^{\dagger}$\footnote{***}
{ e-mail:yoneyama@math.ms.saga-u.ac.jp}}
\bigskip
\centerline{Department of Physics}
\centerline{Kyushu University}
\centerline{Fukuoka, 812 JAPAN}
\bigskip
\centerline{$\dagger$ Department of Physics}
\centerline{Saga University}
\centerline{Saga, 840 JAPAN}
\vskip1cm
\centerline{{\bf Abstracts}}
The two dimensional $CP^1$ model with $\theta$ term is simulated.
We compute the topological charge distribution $P(Q)$ by employing the
``set method" and  ``trial
 function method", which  are effective  in  the calculations for
very  wide range of $Q$ and  large volume.
The distribution  $P(Q)$ shows the Gaussian behavior in the small
 $\beta$ (inverse coupling
 constant) region and   deviates from it in the large $\beta$ region.
The free energy and its moment are calculated as a function of $\theta$.
{}For small $\beta$, the partition function is given by the elliptic theta
function, and the distribution of its zeros
on the complex $\theta$ plane leads to the
 first order phase transition at $\theta=\pi$.
 In the large $\beta$ region, on the other hand, this  first order phase
transition disappears,
 but definite  conclusion concerning the transition is not  reached due to
large errors.
\par
\Date{}
\newsec{Introduction}
The two dimensional $CP^{N-1}$ model is a suitable laboratory to study
dynamics of QCD. Topology is expected to play an important role  in
 non-perturbative nature of the dynamics of such theories.
Recently numerical study of the topological aspects of the  $CP^{N-1}$ model
has made much  progress
\ref\Pisa{A.~Di Giacomo, F.~Farchioni, A.~Papa and E.~Vicari, \PRD{46}, 4630
(1992).}
\ref\Pisb{M.~Campostrini, P.~Rossi and E.~Vicari, \PRD{46}, 2647 (1992).}.
However a full understanding of the dynamics of the model requires study of
 an  additional contribution of  the imaginary part of the action, i.e.,
$\theta$ term.
The  degeneracy of the different topological sector is resolved into a unique
 vacuum labeled by a parameter $\theta$.
 As shown by some   analytic  studies,
various  models with the $\theta$ term, in general,   exhibit a rich phase
structure
\ref\W{U.~-J.~Wiese, \NP{318}, 153 (1989).}
\ref\HIY{A.~S.~Hassan, M.~Imachi and H.~Yoneyama, \PTP{93}, 161 (1995).}
\ref\CR{ J.~L.~Cardy and E.~Rabinovici, \NP{205[FS5]}, 1  (1982).}
\ref\C{J.~L.~Cardy, \NP{205[FS5]}, 17 (1982).}.
It is then worthwhile to study  effects of the $\theta$ term
to  the  $CP^{N-1}$ model \ref\SC{G.~Schierholz,`` $\theta$ Vacua,
 Confinement and The Continuum Limit",
 preprint DESY 94-229, HLRZ 94-63, hep-lat/9412083.}.
{} From a realistic point of view also, it is significant   to clarify the
matter of  strong CP violation in QCD.\par
Introduction of the $\theta$ term does not allow ordinary simulations
because of  the complex Boltzmann factor.
An idea to circumvent the problem is to introduce the constrained updating
of the fields, in which   the topological charge,
being a functional of the dynamical fields, is constrained to take a given
  value $Q$. So the phase factor $e^{i \theta Q}$ is factored out  so that
the partition function  is given by the summation of the probability
distribution $P(Q)$ weighted by $e^{i \theta Q}$  over  all possible  values
  of the topological charge   $Q$.
This algorithm was   adopted in simulating   the two dimensional U(1)
 gauge model $\W$.\par
So far  topological aspects of the  $CP^1$ model has been studied
 considerably well both theoretically and numerically.
Most   works are, however,   limited to the theory without the $\theta$ term.
This is one of our motives for  studying effects of $\theta$ term on
the   model by means of Monte Carlo simulations.
We present here the results of $P(Q)$  and the free energy $F(\theta)$ and
 its moments  as a function of $\theta$ by surveying   comparatively
 wide range of $Q$. Herein the two techniques are involved; one is
 to take the set method $\ref\BBCS{G.~Bhanot, S.~Black, P.~Carter and
R.~Salvador,
 \PL{183}, 331 (1987).}$, and the other is to update by modifying the action to
 the effective  action  by introducing trial probability distributions in
each set. These enable one to reach  very large  $Q$'s.
In ref. \ref\BDSL{G.~Bhanot, R.~Dashen, N.~Seiberg and H.~Levine, \PRL{53},
 519 (1984).}, the model was investigated, and the nature of
the dilute gas approximation was clarified. In the present paper, we are
 interested in
the phase  structure in  $\theta$ and $\beta$ ( inverse coupling constant )
 space. We do simulations   extensively
for  various $\beta$  in   larger
 volume $V$  and wider range of $Q$  using the techniques.\par
{}From a viewpoint of  condensed matter physics as well, the phase structure
of the  $CP^1$ model is worthwhile to study.
The antiferromagnetic quantum  Heisenberg chain with spin $s$ is,
in the large spin limit, mapped to
the two dimensional O(3) non-linear sigma model ($CP^1$ model),
 as an effective theory describing the low energy dynamics.
 The topological nature appears through  the  $\theta$ term
 with $\theta =2\pi s$
\ref\Af{I.~Affleck, ``Field theory methods and  quantum  critical phenomena",
 Proceeding of Les Houches Summer School 1988 Session XLIX,  563  (1990). },
 and then its effect   is expected to
distinguish the dynamics between  the $s=$ integer and half-odd integer cases.
 This feature  is stated as  the  Haldane conjecture that
the $s=$ integer antiferromagnetic Heisenberg  chain develops gap, while
$s=$ half-odd integer one is gapless
\ref\H{F.~D.~M.~Haldane, \PRL{50}, 1153 (1983).}.\par
The above mapping  is based on spin wave approximation in the large $s$ limit.
 However the approximation
 is believed to be good   even for very small $s$.
 It is well known that $s=1/2$ antiferromagnetic Heisenberg model
 is gapless, and correspondingly  the non-linear sigma model with
$\theta =\pi$ would be  critical.
  There have been analytic arguments assuring  this
\ref\SR{R.~Shankar  and N.~Read, \NP{336}, 457 (1990).}.
So far, however,  only a few works have been done in terms of
 numerical calculations \ref\BPW{W.~Bietenholz, A.~Pochinsky and U.~-J.~Wiese,
``Meron-Cluster Simulation of the $\theta$-Vacuum in the $2-d$ O(3)-Model",
preprint CTP 2433, hep-lat/9505019.}.
The present paper also  concerns   this issue  numerically.\par
  As shown in this paper, $P(Q)$ shows qualitatively quite different behavior
for  small and large coupling constants;   it shows clearly  the Gaussian
behavior
 in the small $\beta$ (inverse coupling constant)  region.
We will discuss about the possible first order
 phase transition deducing from the Gaussian distribution.
It is based upon  analytical discussion by using the third elliptic theta
function and the Poisson sum formula.
In the large $\beta$ region, $P(Q)$    systematically deviates
 from the Gaussian.
 We show the structural difference of $F(\theta)$ and
 its moments from the small  $\beta$ region at finite $\theta$.
Near $\theta = \pi$, however, we are  not able to draw a definite conclusion
  about the phase structure  due to the large errors.
 We will discuss this matter in detail.\par
 In the following section we fix the notations and present a brief
account of the algorithm of the simulations.
In section 3 we give  the results.
 In section 4, the partition function zeros are discussed.
  Conclusions and discussion are presented in section 5.\par
\newsec{$CP^1$ model in two dimensions}
\subsec{notations}
We consider the $CP^1$  model with $\theta$ term  on a two space-time
dimensional euclidean lattice  defined by the action
\eqn\act{\eqalign{
 S_\theta= & S -  i \theta \hat{Q},  \cr
 S= & -\beta \sum_{n,\mu} \mid  {\overline z_{n+\mu}} z_n\mid ^2,  \cr
}}
where $\hat{Q}$ is a topological charge, and
 $z_{\alpha,n}$ is a two component complex scalar field ($\alpha=1,2$)
at site $n$ constrained by
$$ {\overline z_n} z_n = \sum_\alpha {\overline z_{\alpha,n}} z_{\alpha,n} =1
$$
and  couples with  the one $\overline z_{n+\mu}$
($\overline z$ is complex conjugation  of $z$) at the nearest
neighbor sites  $n+\mu (\mu=1,2)$.
 \par
The partition function as a function of the coupling constant $\beta$ and
 $\theta$ is  defined  by
\eqn\part{ Z(\theta)=\int \prod_n dz_n d{\overline z_n}  e^{-S_\theta}/
\int \prod_n dz_n d{\overline z_n}  e^{-S}
}
 and the free energy density $F(\theta)$ is  given by
\eqn\fre{
F(\theta)=-{1 \over V }\log Z(\theta),
}
where $V$ is  volume of the system.
The   topological charge $\hat{Q}$ is the number of times the fields
 cover the sphere $S^2$.  The lattice counterpart we adopt is that of the
geometrical  definition in ref. \ref\BL
{B.~Berg and M.~L\"uscher, \NP{190[FS3]}, 412 (1981). };
the charge density  $\hat{Q}(n^*)$ at dual site $n^*$ is given by
\eqnn\tpc$$\eqalignno{
\hat{Q}(n^*)={1 \over 2\pi }{\rm Im} \Bigl\{ &\ln
\big[{\rm Tr} P(n)P(n+1)(P(n+1+2)\big] \cr &+
 \ln \big[{\rm Tr }P(n)P(n+1+2)P(n+2)\big]\Bigr\},&\tpc\cr}$$
where $P(n)_{\alpha\beta} = z_{\alpha n} {\overline z_{\beta n}}$ and
$n$ is the left corner of the plaquette with center $n^*$.
This amounts, in terms of $z$,  to the topological charge
\eqn\tpcc{
\hat{Q}={1 \over 4\pi } \sum_{n,\mu,\nu} \epsilon_{\mu \nu}
 ( \theta_{n,\mu} +  \theta_{n+\mu,\nu} -
\theta_{n+\nu,\mu} - \theta_{n,\nu} ),
}
where $\theta_{n,\mu}= {\rm arg} \{ {\overline z_n}z_{n+\mu} \}$.\par
In order to simulate the model with the complex Boltzmann factor,
we follow the Wiese's idea \W.
It,  in principle, introduces the constrained updating
of the fields, in which   the topological charge,
being a functional of the dynamical fields, is constrained to take a given
  value $Q$.
So the phase  factor $e^{i \theta Q}$ is factored out, so that
 the partition function  is given by the summation
of the probability distribution $P(Q)$ weighted by $e^{i \theta Q}$
in each  $Q$ sector.
This amounts, in practice,  to    calculate first the probability
 distribution  $P(Q)$ at $\theta=0$  and to be followed by taking  the
Fourier  transform of $P(Q)$ to get the partition function $Z(\theta)$ as
\eqn\four{
Z(\theta)=\sum_Q  P(Q) e^{ i \theta Q},
}
where $P(Q)$ is
\eqn\tcd{
 P(Q)={\int \prod_n dz_n d{\overline z_n} ^{(Q)} e^{-S}
 \over \int \prod_n dz_n d{\overline z_n} e^{-S} }.
}
Here $\prod_n dz_n d{\overline z_n} ^{(Q)} =  \prod_n dz_n
d{\overline z_n} \delta_{\hat{Q},Q}$, i.e. ,  the integration measure
 restricted to the configurations with given  $Q$, where $\delta_{\hat{Q},Q}$
is the
Kronecker's delta. Note that $\sum_Q P(Q)=1$.
Expectation value of an observable $O$ is given in terms of $P(Q)$ as
\eqn\evq{
\langle O \rangle _{\theta} = {\sum_Q P(Q) \langle O \rangle_Q
e^{ i \theta Q}
\over
\sum_Q P(Q) e^{i \theta Q} },
}
where $\langle O \rangle_Q$ is the expectation value of $O$  at $\theta=0$
 for a given $Q$ sector
\eqn\epv{
\langle O \rangle _Q ={\int  \prod_n dz_n d{\overline z_n}^{(Q)} O e^{-S}
\over
 \int  \prod_n dz_n d{\overline z_n}^{(Q)}  e^{-S} }.
}
\smallskip
\subsec{algorithm}
We measure the topological charge distribution $P(Q)$ by Monte Carlo
simulation by the Boltzmann weight  $\exp(-S)$, where $S$ is defined
by \act.
The standard Metropolis method is used to
update  configurations. To calculate $P(Q)$ effectively, we apply
(i) the set method and (ii) the trial distribution method simultaneously.
In the following, we explain briefly the algorithm to make the paper
self-contained.
All we have to calculate $P(Q)$  is to count how many
times the configuration of $Q$ is visited by the histogram method.
The distribution $P(Q)$ could damp very rapidly as $\vert Q \vert$  becomes
large.
We need to calculate the $P(Q)$ at large  $\vert Q \vert$'s which would
contribute to $F(\theta)$, $\langle Q \rangle_\theta$ and
$\langle Q^2 \rangle_\theta$ because
they are obtained by the Fourier transformation of $P(Q)$
and its derivatives.
 Further, the error of $P(Q)$ at large
$\vert Q \vert$ must be suppressed as small as possible. These are reasons
 why  we apply two techniques mentioned above.
Since $P(Q)$ is analytically shown to be even function and is certified by
simulation, we restrict to the range of $Q$ to $\geq 0$.\par
The range of $Q$ is grouped into sets $S_i$ ; $S_1 (Q=0 \sim 3)$,
$S_2 (Q=3 \sim 6)$, $\cdots$ , $S_i (Q=3(i-1) \sim 3i)$, $\cdots$  (set
method).
Monte Carlo updatings  are   done as follows by starting from a
configuration within a fixed  set $S_i$.
 When  $Q$ of a trial  configuration ${C_t}$
 stays in  one of the bins within  $S_i$, the  configuration ${C_t}$
is accepted, and the count of the  corresponding $Q$ value is increased by one,
while  when  ${C_t}$ goes out of the set $S_i$, ${C_t}$ is rejected,
 and the count of $Q$ value of the old configuration is increased by one.
This is done for all sets $S_i$ ; $i$ = 1, 2, $\cdots$.\par
Another of the two techniques is to  modify the Boltzmann weight by introducing
 trial  distributions $P_t(Q)$ for each set (trial distribution method).
 This is to remedy $P(Q)$ which falls  too rapidly even within a set
 in some cases.
We make the counts at
$Q=3(i-1), 3(i-1)+1, 3(i-1)+2$ and $3i$ in each set $S_i$ almost the same.
 As the trial distributions $P_t(Q)$'s, we apply the form
$$
P_t(Q) = A_i \exp [ - \bigl( C_i(\beta) / V \bigr) Q^2 ],
$$
where the value of $C_i(\beta)$ depends on the set $S_i$,
and $A_i$ is a constant.
That is,  the  action during  updatings is modified to the effective one
such as $S_{\rm eff}=S + \log P_t(Q) $.
 We adjust
$C_i(\beta)$ from short runs to get almost flat distribution
at every $Q$ in  $S_i$. \par
To obtain the normalized distribution $P(Q)$ in the whole range of $Q$
from the counts at each set, we make matchings  as follows:
\itemitem{i).} At each set $S_i$ ( $i$ = 1, 2, $\cdots$ ), the number of counts
is multiplied
by $P_t(Q)$ at each $Q$. We call the multiplied value $N_i(Q)$,
which is hopefully proportional to the desired topological charge distribution
$P(Q)$.
\itemitem{ii).} In order to match the values  in two neighboring sets $S_i$ and
$S_{i+1}$,
we rescale $N_{i+1}(Q)$ so that $N_{i+1}(Q) \rightarrow N_{i+1}(Q) \times r$,
where $r$ =$N_i(Q=3i) / N_{i+1}(Q=3i)$,  the ratio of the number of counts at
the right edge of $S_i$ to that at the left edge of $S_{i+1}$.
 These manipulations are performed over all the sets.
\itemitem{iii).}The rescaled $N_i(Q)$'s are  normalized to
 obtain  $P(Q)$ such that
$$
P(Q) = { N_i(Q) \over \sum_i\sum_Q N_i(Q) }.
$$
\bigskip
\newsec{Numerical Results}
   We use square lattices with the periodic boundary conditions.
 Lattice sizes are $V = L \times L$, and $L$ ranges from  $L =$ 24, 36, 48  to
 72. The total number of  counts in each set is $10^4$. The error analysis is
discussed in Appendix.
To check the algorithm, we calculated the internal energy. It agrees with
the analytical results of  the strong and weak coupling expansions \BL.
Using the calculated $P(Q)$, we will estimate the free energy $F(\theta)$
and its derivative $\langle Q \rangle_\theta$, respectively.
\smallskip
\subsec{topological charge distribution $P(Q)$}
In this subsection we discuss the topological charge distribution $P(Q)$.
Partition function can be given by the measured $P(Q)$ as in $\four$
  in principle, but
we should be careful for estimating $Z(\theta)$ from $P(Q)$.
Since $P(Q)$ is very sharply decreasing function of $Q$, its Fourier series
$Z(\theta)$ is
drastically affected by statistical fluctuations of $P(Q)$.
For example, consider two different $Q$ values, say, $Q_1$ and $Q_2$
 ($Q_1 \ll Q_2$).
Small error $\delta P(Q_1)$ at $Q_1$ could cause very large effects
 to $Z(\theta)$ because $P(Q_2)$ itself is sometimes much smaller than $\delta
P(Q_1)$.
So the effort to obtain $P(Q)$ at large $Q$ may be useless if we allow these
fluctuations
at the small value of $Q$. In order to avoid this problem, we first fit the
measured $P(Q)$ by the
appropriate functions $P_{\rm fit}(Q)$ and obtain $Z(\theta)$ using Fourier
transforming
from $P_{\rm fit}(Q)$.  We apply the chi-square-fitting
to the logarithm of the measured $P(Q)$ in the form of
 polynomial functions of $Q$
$$
P(Q) = \exp \Big[ \sum_n a_n Q^n \Big].
$$
In the following, we present the results of  $\beta$ and  volume dependence of
 $P(Q)$ .
 \par
In Fig.1, we show the measured $P(Q)$ for various $\beta$'s ($\beta =$ 0.0,
0.5, $\cdots$, 3.5)
 for a fixed volume ($L=24$). As $\beta$ varies, $P(Q)$ smoothly changes from
strong to weak coupling regions.   In the strong coupling regions $(\beta \lsim
2.0)$,
 $P(Q)$ shows Gaussian behavior. In the weak coupling regions
 $(2.75 \lsim \beta)$,
$P(Q)$ deviates gradually from the Gaussian form, being enhanced at large $Q$
compared to the Gaussian. In order to investigate  the difference between the
two regions
 in detail, we use the chi-square-fitting to $\log P(Q)$. Table I shows the
results of the fittings, i.e.,
the coefficients $a_n$ of the used  polynomial $\sum_n a_n Q^n$  for various
$\beta$'s with
 the resulting $\chi^2/d.o.f$.  (i) For $\beta \lsim 2.0$, $P(Q)$'s are
 indeed fitted well by the Gaussian
form. (ii) For $\beta \gsim 2.75$,  terms up to   quartic one
 are    needed for sufficiently good fitting. The linear term, in particular,
is important for
fitting the data at very small $Q$ values.
The value $Q_{\rm Max}$, which is the largest $Q$ of the range in
consideration, is also shown
in the table. It is chosen so that the ratio $P(Q_{\rm Max}) / P(0) \approx
10^{-20}$ in the weak couplings.
(iii) Between the strong and weak couplings ($2.0\lsim\beta\lsim 2.75$) the
fittings according to the
polynomial turn out to be  very poor ( $\chi^2/d.o.f. \approx 250 $ ).
 It may  indicate   the existence of  a transitive region between the Gaussian
and non-Gaussian
regions.
(iv) Apart from this region,   each of the coefficients  change smoothly
from the strong to weak coupling
 regions as shown in Table I.\par
Here we discuss the volume dependence.
In the strong coupling regions, $P(Q)$ is  fitted very well by Gaussian for all
values
 of $V$
$$
P(Q) \propto \exp \left( - \kappa_V(\beta) Q^2 \right).
$$
where the coefficient $\kappa_V(\beta) (=a_2)$ depends on $\beta$ and $V$.
Fig.2 shows $\log \kappa_V(\beta)$  vs. $\log V$ for a fixed $\beta(=0.5)$.
We see  that  $\kappa_V(\beta)$ is clearly  proportional to $1/V$
$$
\kappa_V(\beta) = C(\beta)/V.
$$
This $1/V$-dependence  of the Gaussian  behavior    determines the phase
structure of
the strong coupling region.
This will be discussed in detail  numerically  in \S 3.2 and
analytically  in \S 4.\par
 The proportionality  constant $C$ depends  on $\beta$.
As $\beta$ becomes large, $C(\beta)$ monotonically increases;
$C(\beta=0.0)=10.6$, $C(\beta=0.5)=12.3$, $C(\beta=1.0)=15.5$.
 \par
Fig.3. shows the volume-dependence of $P(Q)$ for $L= 24$, 36, 48, and 72
 in the weak coupling regions ($\beta=3.0$).
We do not find  the $1/V$-law  as in  the strong coupling regions, but
 a  clear volume dependence is observed. It causes the different behavior of
$F(\theta)$ from that in the strong coupling regions.
\par
%
%
\bigskip
\subsec{Free energy and expectation value of topological charge }
\smallskip
Partition function $Z(\theta)$ as a function of $\theta$ is given
 by $\four$ from $P(Q)$.
The free energy is
\eqn\f{
F(\theta) = -{1 \over V} \log Z(\theta).
}
In general, the $n$-th order of the moment is given by the derivatives of
$F(\theta)$
\eqn\q{
\langle Q^n \rangle_\theta = -(-i)^n {d^n F(\theta) \over d\theta^n.}
}
\par
%
%
In the strong coupling region, we have seen  the Gaussian behavior of $P(Q)$,
and
the  $1/V$-law  appears to hold   up to $L=72$.
It is natural  to expect  that this behavior persists to $V \rightarrow
\infty$.
Let us look at how  the $1/V$-law affects  $F(\theta)$ and
$\langle Q \rangle_\theta$.
By putting $C(\beta)=12.3$ in
$P(Q) \propto \exp \left[ - \left( C(\beta)/V \right) Q^2 \right] $
for $\beta=0.5$,
 we calculate   $F(\theta)$ and  $\langle Q \rangle_\theta$ from $\f$ and $\q$.
Fig's 4 and 5 show their  volume dependence. As $V$ is increased, $F(\theta)$
 very  rapidly (already at  $L=6$) approaches the quadratic form in $\theta$
from below.
Its first moment $\langle Q \rangle_\theta$ develops a peak near
$\theta=\pi$, and the position of the peak quickly approaches $\pi$ as $V$
increases.
The jump in $\langle Q \rangle_\theta$ would arise at $\theta = \pi$
as $V \rightarrow \infty$.
It indicates the first order phase transition at $\theta=\pi$.\par
In the weak coupling regions, on the other hand, we see the different behavior.
Fig.6 shows  $F(\theta)$ at $\beta=3.0$.
For $\theta \lsim \pi/2$, $F(\theta)$ is volume independent, while
for $\theta \gsim \pi/2$, the clear volume dependence appears, where
 $F(\theta)$ decreases as $V \rightarrow $ large unlike in the strong coupling
case.
We have checked that the result for $L=20$ agrees with that in ref.$\BDSL$
 within errors.
The expectation value $\langle Q \rangle_\theta$ is shown in Fig.7.
The singular behavior at $\theta=\pi$, which was seen
in the small $\beta$ region, disappears.  The peak gets round and
its locus moves towards small $\theta$ as $V$ increases,
which is opposite to Fig.5. \par
We should make a remark about the errors in the figures.
As a general tendency,
larger errors arise for
larger volume and/or for  $\theta \approx \pi$.
It is associated with the algorithm to calculate $Z(\theta)$, $\four$, in which
$e^{i \theta Q} \approx (-1)^Q $ for $\theta=\pi$ yields large cancellation
for slowly falling $P(Q)$ (the  behavior at large $V$) in the summation.
It causes large errors of the observables due to the denominator in $\evq$.
 This is just the same as   the    so called  sign problem
\ref\LGSWSS{E.~Y.~Loh~Jr., J.~E. Gubernatis, R.~T. Scalettar, S.~R.~White,
D.~J.~Scalapino
 and R.~L.~Sugar, \PRB{41}, 9301 (1990).}
 which is notorious
 in the quantum Monte Carlo simulations applied to systems of strongly
correlated electrons.
 \par
%
\newsec{Gaussian distribution and the partition function zeros }
In the previous sections, we have seen that $P(Q)$ is Gaussian in
 small  $\beta$ region.
In this section we shall look into the detail  of its consequence by
paying attention to  the partition function zeros in the complex $\zeta$
  plane ($\zeta=e^{i  \theta}$).
Study of the  partition function zeros is regarded as an alternative
to investigate the critical phenomena.  The zeros accumulate in infinite
volume limit to the critical point, and how fast they approach the point as $V$
increases tells the order of the phase transition
\ref\IPZ
{C.~Itzykson, R.~B.~Pearson and J.~B.~Zuber \NP{220[FS8]}, 415 (1983).}
\ref\FB{M.~E.~Fisher and A.~N.~Berker, \PRB{26},  2507 (1982).}.
If  Gaussian behavior $P(Q) \propto \exp [- \left( C(\beta)/V \right) Q^2]$
persists to
 infinite volume limit,  the partition function
is expressed by  the third  elliptic theta function
%
\eqn\ell{
\vartheta_3(\nu,\tau)=\sum_{Q=-\infty}^\infty p^{Q^2} \zeta^{Q}
}
as
$$
Z(\theta) \propto \vartheta_3(\nu,\tau),
$$
where  $p=\exp[-C(\beta)/V] \equiv \exp(i \pi \tau)$
  and $\zeta=e^{i\theta} \equiv \exp(i 2 \pi \nu)$.
In order to look for the partition function zeros in the
complex $\zeta$ plane, it is convenient
 to use infinite product expansion of $\vartheta_3$
\eqn\ipe{
\vartheta_3(\nu,\tau)=\prod_{m=1}^\infty (1-p^{2 m})
\prod_{n=1}^\infty [(1+p^{2 n-1}\zeta)(1+p^{2 n-1}\zeta^{-1})].
}
Zeros of $Z(\theta)$ are all   found  easily on the negative real axis of the
complex $\zeta$ plane as
\eqn\zer{
 \zeta=-e^{-(2 n-1)C/V}, -e^{+(2 n-1)C/V}
}
for $n$ = 1, 2, $\cdots$, $\infty$.
In the complex $\theta$ plane, equivalently,  these zeros  are located at
$$
\theta=\pi\pm i (2 n-1) C/V.
$$
It thus follows that the $1/V$-law  approaching  the  critical point
 $\theta_c=\pi$  indicates the first order phase transition $\IPZ$ $\FB$.\par
 An alternative to the above way of looking is to use the
Poisson  sum formula to the sum \four.
\eqn\poi{Z(\theta) \propto \sum_{Q=-\infty}^\infty e^{-C Q^2/V} e^{i\theta Q}
   =\sqrt{V\pi/C} \sum_{n=-\infty}^\infty e^{-(\theta -2\pi n)^2 V/4 C}.
}
For  $V \gg 1$ and near $\theta=\pi$, the sum on the right is well approximated
 by  two terms ( $n=0$ and 1 ),
\eqn\pois{ (4.4)
\approx\sqrt{V\pi/C}\left[ e^{-\theta^2 V /4 C}
  + e^{-(\theta-2\pi)^2 V/4 C} \right].
}
It follows that   the partition function has infinite zeros at
 $\theta=\pi + i (2 n+1)C/V$, where $n$ is integer.
 Again the  $1/V$-law  means the existence of the first order phase transition.
 This result is  in complete agreement with that from $\vartheta_3$ function
discussed above.
  To see to what extent the approximation $\pois$ is good, we compare the
 resulting  $F(\theta)$ and $\langle Q \rangle_\theta$  from $\pois$ with
those of
Monte Carlo simulations. They agree each other.
\bigskip
\newsec{Conclusions and discussion}
  We have seen  that $P(Q)$ is Gaussian in the small $\beta$ region.
As shown in the last section, it leads to the first order phase
 transition. This behavior is very much like  the   $d=2$ U(1) gauge model
 with $\theta$ term \W, where $P(Q)$  is Gaussian for all values of
 the coupling constant \ref \HITY{A.~S.~Hassan, M.~Imachi, N.~Tsuzuki and
H.~Yoneyama,
 ``Character Expansion, Zeros of partition function and
$\theta$ term in U(1) gauge theory", preprint KYUSHU-HET-25, SAGA-HE-86,
 hep-lat/9508011.}.   There the analytic form of $P(Q)$ is given.
It may  also  be interesting to study the $CP^1$ model from  the
 renormalization group point of view, which might show the singular
behaviors of the renormalization group flows similar to the U(1) case \HIY.
 \par
 In large $\beta$ region, on the other hand,  $P(Q)$ differs  from the
Gaussian behavior. Consequently, the free energy $F(\theta)$ and the moment
$\langle Q \rangle_\theta$ show the quite different behaviors from those in the
small $\beta$
regions. The signal of the first order transition disappears. To understand
those behaviors,
It would be helpful to consider the dilute gas approximation,
where instantons of charge $Q=\pm 1$  are randomly distributed.
Let us assume that the probability distribution  $P_n$ ( $P_{\overline n}$
),  in which $n$ instantons (${\overline n}$ anti-instantons) generate,
obeys the    Poisson distribution
$P_n=\lambda^n e^{-\lambda}/n! ( P_{\overline n}=\lambda^{\overline n}
 e^{-\lambda}/{\overline n}! ) $. The topological charge distribution function
 $P(Q)$ is given by the modified Bessel's function as
 $P(Q)=e^{-\lambda} I_Q(\lambda)$, where $\lambda$ is
average number of instantons (anti-instantons).
For $\lambda \gg 1$, $I_Q(\lambda)$ is approximated by  $\exp(-Q^2/2\lambda)$.
The  $\lambda$ can then  be identified as $V/2C$,
 which is natural since the average number   is proportional to
the volume $V$. As $\beta$ increases, $C(\beta)$ increases
 (section 3),  that is, the average number of instantons decreases;
 as  $\beta \rightarrow \infty$ (zero temperature limit), the configurations
  vary slowly so that the configurations with large $Q$ are unlikely to
contribute to the partition function. In large $\beta$ region, the  behavior of
$I_Q(\lambda)$ as a function of $Q$ is  qualitatively the same with the
 result of the simulations.
 Precisely speaking, however, they are different, and actually the difference
is attributed to
 the asymptotic scaling of the topological susceptibility in ref. \BDSL.
\par
It is expected from the Haldane conjecture
that the second order phase transition would occur at $\theta=\pi$.
We, however,  seem to fail confirming it. The first order phase transition  in
small $\beta$ region would have to mutate to the second order one at some
$\beta$,
if it occurred.
In the large $\beta$ region,  as discussed in section 3, the volume dependence
of the
 results is large in the interesting  region of $\theta$, and the statistical
errors
 mask the  nature.
For $L=72$, the maximal lattice extension of our study, $F(\theta)$ still
changes considerably and
gets very large errors   for $\theta \gsim \pi/2$.  Consequently, so do  its
moments  for
  a  wider range of $\theta$.
This is due to the large correlation length in the large  $\beta$ region, and
 the finite size effect is not negligible.
The large fluctuations   come  from  the same origin as the so called
sign problem \LGSWSS, which arises in the strongly correlated electronic system
in  the condensed matter physics.
In order to circumvent the problem, we must address the issue of the lattice
effect.
It is worthwhile  to pursue the issue  treated in the present paper
from the the improved point of view
 such as the  perfect action
$\ref\HN{P.~Hasenfratz and F.~Niedermayer, \NP{414}, 785 (1994).}$
$\ref\DFP{M.~D'Elia, F.~Farchioni and A.~Papa,
``Scaling and topology in the 2-$d$ O(3)-$\sigma$ model on the lattice
 with the perfect action", preprint IFUP-TH 23/95, hep-lat/9505004.}$.
Recently, the second order phase transition has been found numerically by
formulating
the model in terms of clusters with fractional topological charge $\pm 1/2$
\BPW.
\par
Some numerical studies of the $CP^{N-1}$ model with $N > 2$
have been done without $\Pisa$ and with the $\theta$ term \SC.
In the latter case for $CP^3$, interestingly, the first order transition is
 observed at finite $\theta$ which is smaller than $\pi$ \SC. \par
\vskip1cm
\centerline {\bf Acknowledgment}
We are grateful to the colleagues for useful discussion.
We also wish to thank S.~Tominaga for discussion on the algorithm.
The numerical simulations were performed on the computer Facom M-1800/20 at
RCNP,
Osaka University.
This work is supported in part by a Grant-in-Aid for Scientific Research
from the Japanese Ministry of Education, Science and Culture (No.07640417).
One of the authors (A.~S.~H.) is grateful for the scholarship from the
Japanese  Government.
\vfill\eject
\centerline{{\bf Appendix }}
\smallskip
In this appendix, we discuss briefly
the error analysis when ``set method" and ``trial
distribution method" are used.\par
We consider first the simple case where a single set is adopted and,
as trial distribution, $P_t(Q) = 1$.
It is known that the counts in the histogram method essentially
obeys multinomial distribution and
that the error of counts at $Q$ ($count(Q)$) is estimated by the variance of
the
distribution \BBCS.
For each  $Q$, the variance is
$$
\sigma^2(Q) = N \cdot {count(Q) \over N} \left( 1 - {count(Q) \over N} \right),
$$
where $N$ is the total counts. Therefore $P(Q)$ is estimated by
$$
P(Q) = {count(Q) \over N}  \pm \delta P(Q),
$$
where $\delta P(Q) = \sigma (Q) / N$.
The relative error ($\delta P / P$) at large $Q$
is given by
$$
{\delta P(Q) \over P(Q)} \approx {\sigma (Q) \over count(Q) }
= {1 \over \sqrt{ count(Q)} }. \eqno (A.1)
$$
It could become very large at large $Q$ when $P(Q)$ is rapidly decreasing
function
of $Q$.\par
When the above two methods are adopted, the relative error decreases as
follows.
The trial distribution method makes $count(Q)$ almost independent of $Q$.
The variance $\sigma(Q)$ also becomes
 almost constant at each $Q$. Accordingly, $P(Q)$ is given by
$$
P(Q) = P_t(Q) \left( count(Q) \pm \sigma(Q) \right),
$$
which leads to the relative errors at any $Q$
$$
{\delta P(Q) \over P(Q)} = {\sigma  \over count } \approx {\rm constant}.
$$
This is quite  an improvement compared to (A.1).
When the set method is further  used, the constant errors  do not
 propagate over different sets \W.
\vfill\eject
\listrefs
\vfill\eject
\centerline{\bf Table caption}
\bigskip
\item{Table I.}   The results of chi-square-fitting to $\log P(Q)$ in terms of
the
polynomial $\sum_n a_n Q^n $ for various $\beta$.  Fittings are performed to
the data
in the range  from $Q=0 $  to  $Q_{\rm Max}$. The resulting $\chi^2/d.o.f.$'s
are also listed.
For the data $\beta=0.0$, 0.5 and 1.0,  Gaussian fitting is performed.
\par
\vskip 1.5cm
\centerline{\bf Figure captions}
\bigskip
 \item{Figure 1.} The topological charge distribution $P(Q)$ vs. $Q^2$ for
$\beta=0.0$ to 3.5. The lattice size is $L=24$.
The  data only for $Q \leq 21$ are plotted.  The lines are shown for the
guide of eyes.\par
\item{Figure 2.}$\log a_2 ( = \log \kappa_V )$ vs. $\log V$. $\beta = 0.5$.
The $1/V$ behavior is clearly seen. \par
\item{Figure 3.}$P(Q)$ vs. $Q^2$ for $\beta=3.0$.
The lattice size $L$ is taken to be  24, 36, 48 and 72.\par
\item{Figure 4.}Free energy $F(\theta)$  for $\beta=0.5$.
  Lines are shown for $V=$ 16, 25 and 36
 in order from below.\par
\item{Figure 5.}The expectation value of the topological charge $\langle Q
\rangle_\theta $
 for $\beta=0.5$.  Lines are shown for $V=$ 16, 25 and 36
 in order from below.  The peak of the curve becomes sharper quickly as $\theta
\rightarrow \pi$.\par
\item{Figure 6.}$F(\theta)$  for $\beta=3.0$. $L$ is chosen  to be  24
(square),
36 (triangle), and 48 (circle).
Values of $F(\theta)$ are plotted based on the parameters $a_n$ obtained by the
fittings explained in the text. The parameters $a_n$ for $L=24$ are shown in
Table I.
Those for $L=36$ and 48 are obtained in the same process as for $L=24$.
The lines are shown for the guide of eyes.
The   volume dependence appears  clearly at $\theta \gsim \pi/2$. \par
\item{Figure 7.}$\langle Q \rangle_\theta $ for $\beta=3.0$. $L$ is the same as
those
in Fig.6.  Error bars for the data of $L=48$ are not drawn because they are too
 large.\par
\vfill\eject
\eject
$$\vbox{
\vskip 5cm
\epsfysize=.950\hsize
\epsfbox{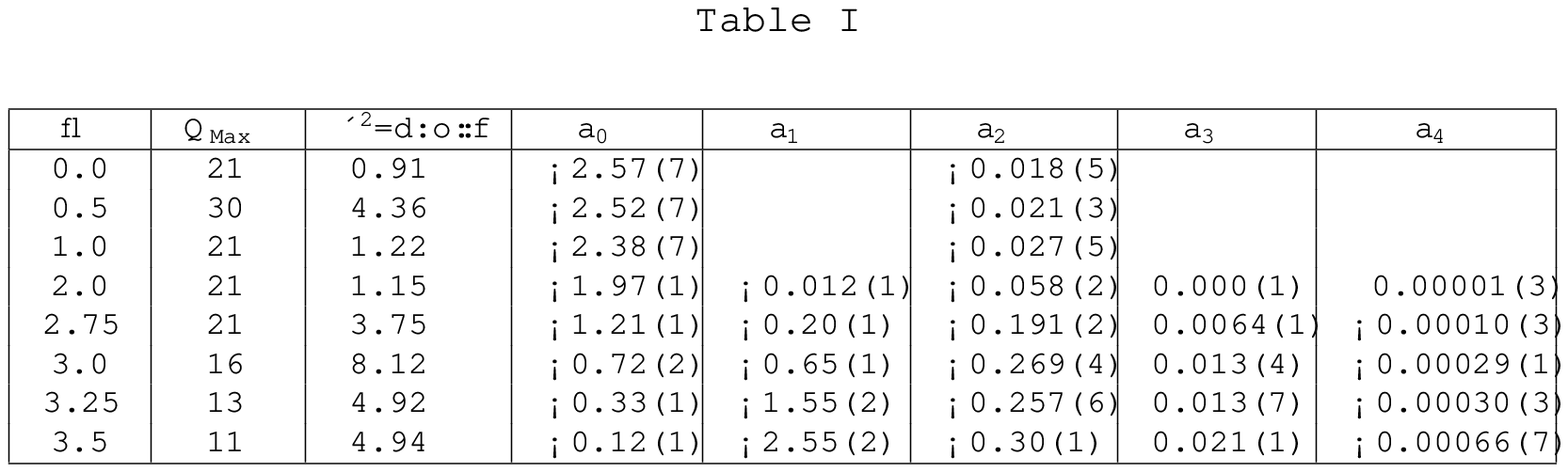}
}$$
\smallskip
\vfill
\eject
$$\vbox{
\vskip 2cm
\centerline{Figure 1}
\bigskip
\epsfysize=1.1\hsize
\hskip.05cm
\epsfbox{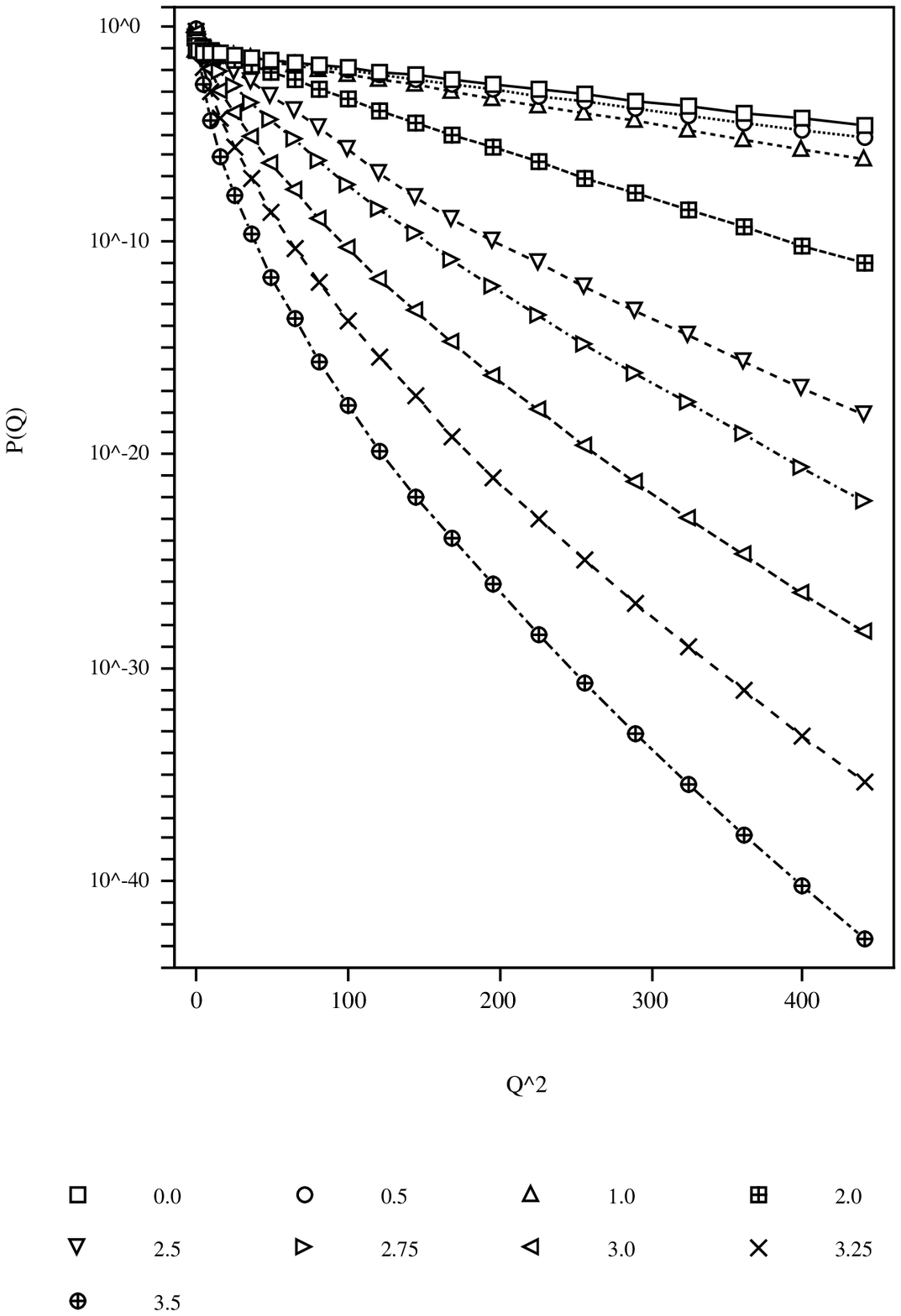}
}$$
\smallskip
\vfill
\eject
$$\vbox{
\vskip 3cm
\centerline{Figure 2}
\bigskip
\hskip1.3cm
\epsfysize=.6\hsize
\epsfbox{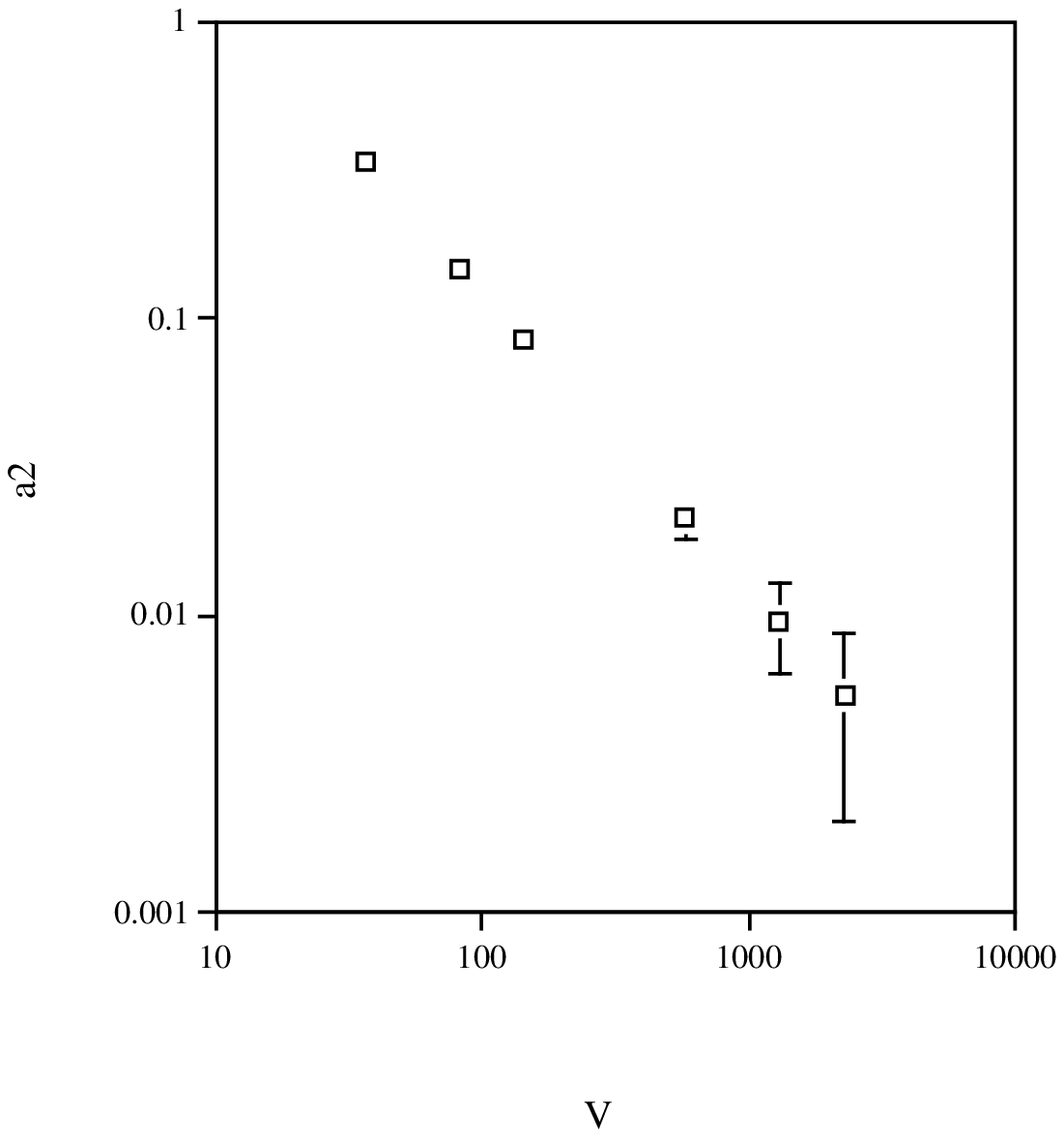}
}$$
\smallskip
\vfill
\eject
$$\vbox{
\vskip 2cm
\centerline{Figure 3}
\bigskip
\epsfysize=1.1\hsize
\hskip.05cm
\epsfbox{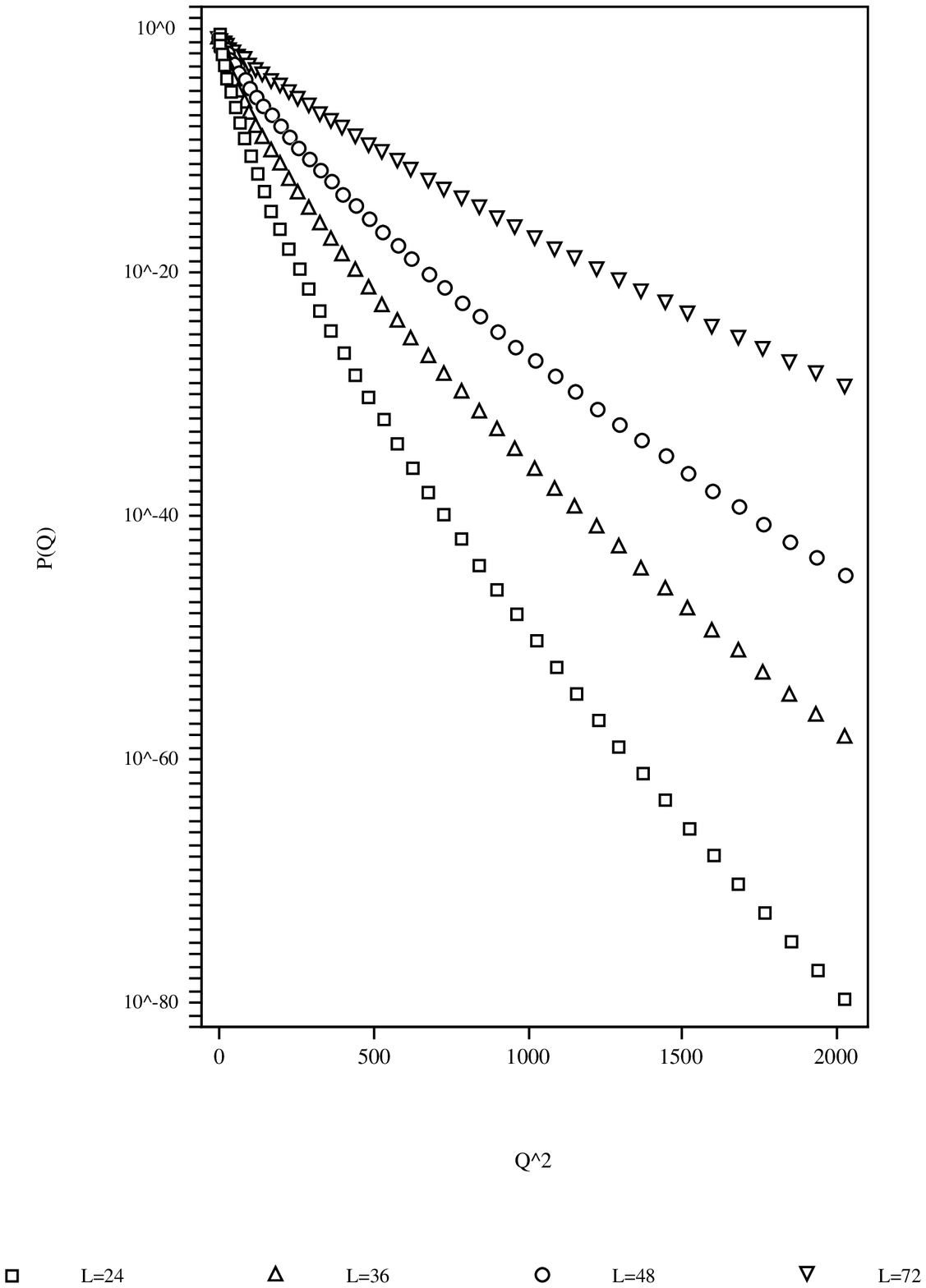}
}$$
\smallskip
\vfill
\eject
$$\vbox{
\vskip 1.7cm
\centerline{Figure 4}
\bigskip
\hskip.2cm
\epsfysize=.5\hsize
\epsfbox{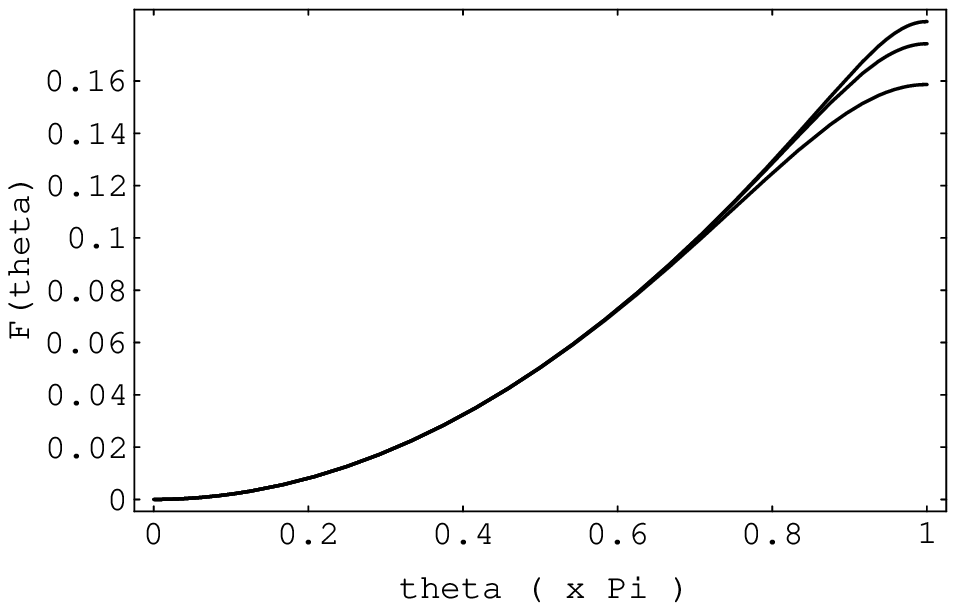}
\vskip 1cm
\centerline{Figure 5}
\bigskip
\hskip.2cm
\epsfysize=.5\hsize
\epsfbox{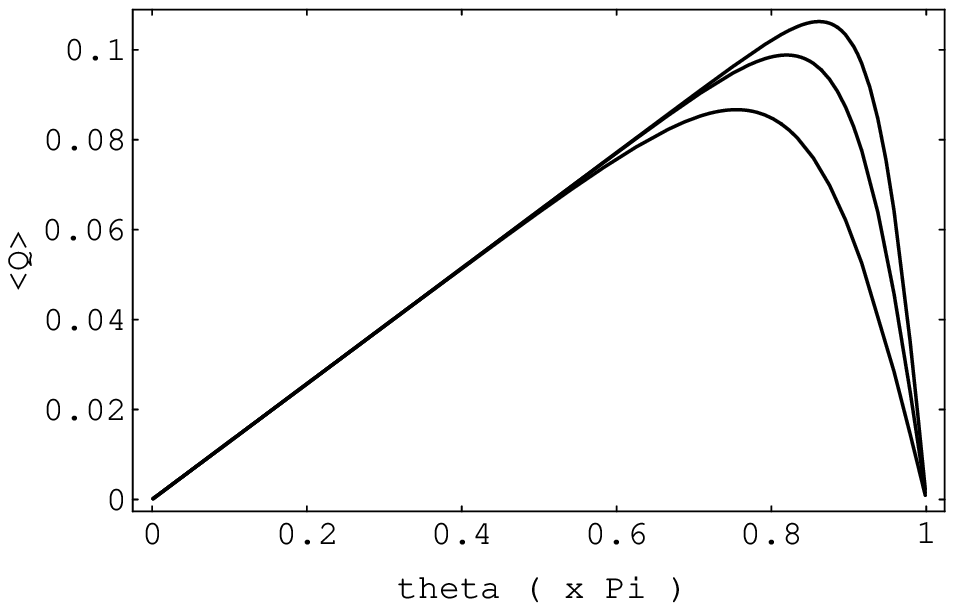}
}$$
\smallskip
\vfill
\eject
$$\vbox{
\vskip 2cm
\centerline{Figure 6}
\bigskip
\epsfysize=.9\hsize
\hskip.05cm
\epsfbox{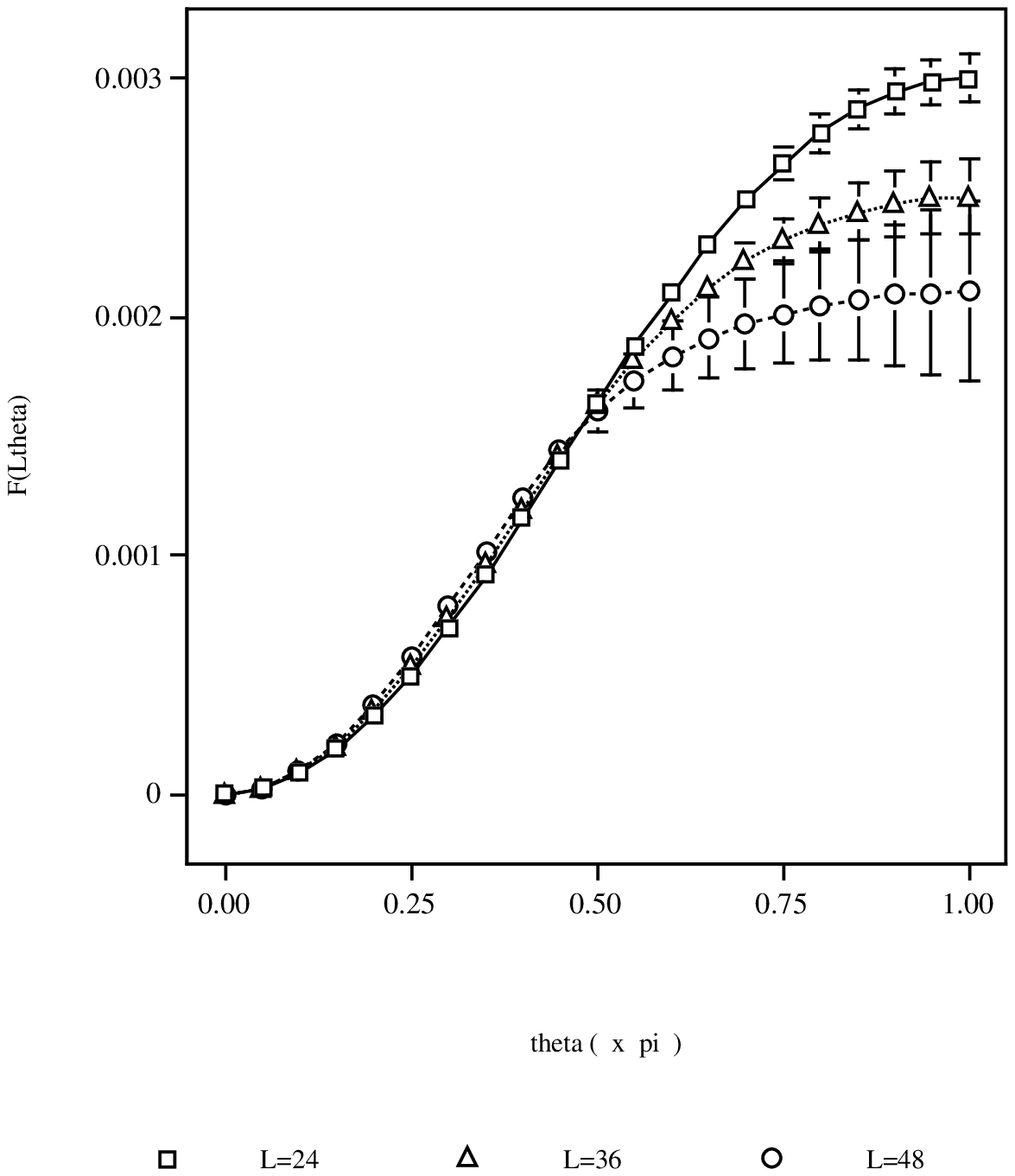}
}$$
\smallskip
\vfill
\eject
$$\vbox{
\vskip 2cm
\centerline{Figure 7}
\bigskip
\epsfysize=.9\hsize
\hskip.05cm
\epsfbox{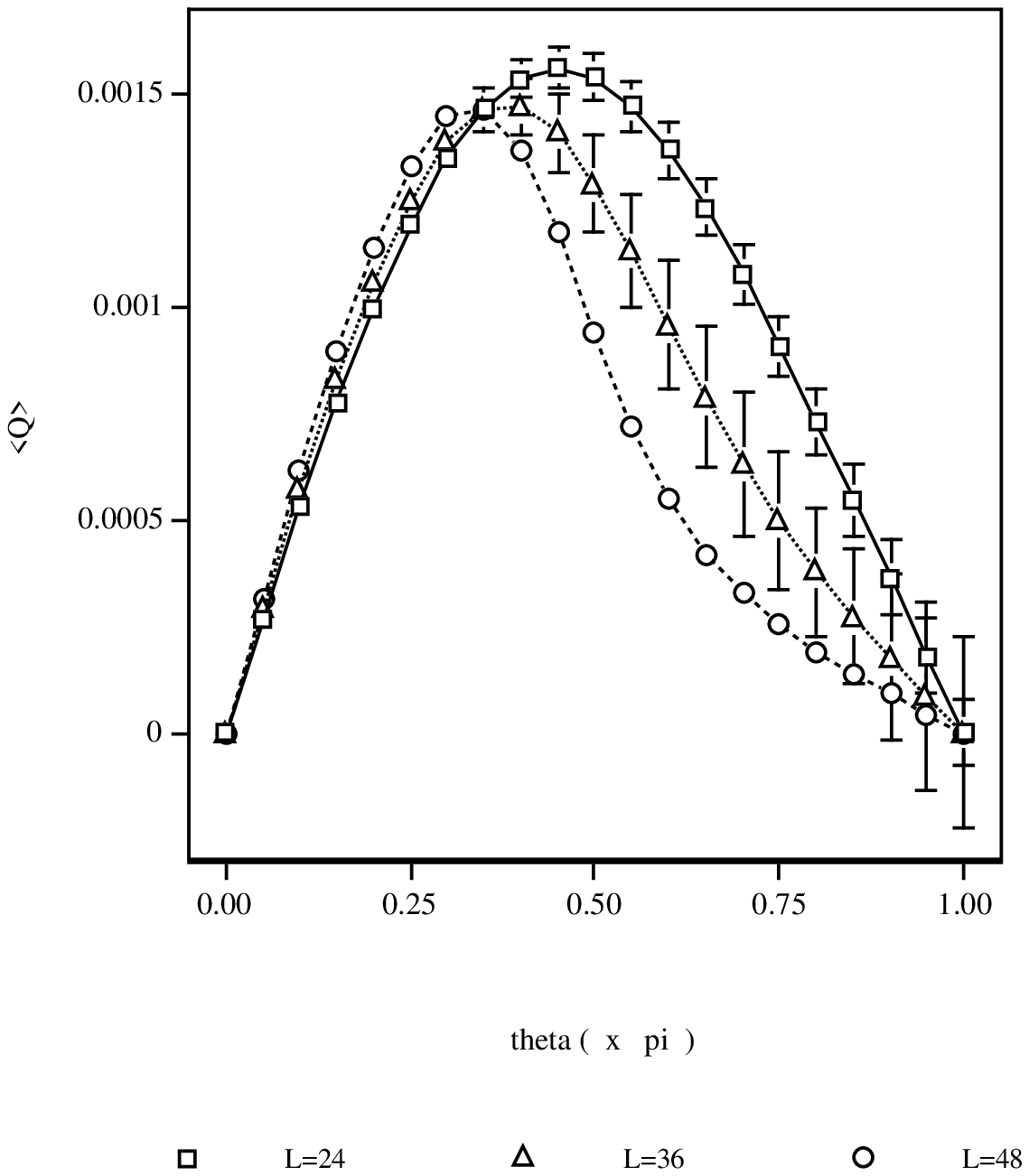}
}$$
\smallskip
\vfill
\eject
\bye